\documentclass[conference,letterpaper,10pt]{IEEEtran}
\usepackage{cite}
\usepackage{amsmath,amssymb,amsfonts}
\usepackage{graphicx}
\usepackage{booktabs,makecell}

\begin{document}

\title{EaaS: A Service-Oriented Edge Computing Framework Towards Distributed Intelligence}
\author{
Mingjin~Zhang,~Jiannong~Cao,~\textit{Fellow,~IEEE,}~Yuvraj~Sahni,~Qianyi~Chen,~Shan~Jiang,~Tao~Wu\\
Department of Computing, The Hong Kong Polytechnic University\\
\{csmzhang, csjcao, cssjiang\}@comp.polyu.edu.hk
}
\maketitle

\thispagestyle{plain}
\pagestyle{plain}

\begin{abstract}
Edge computing has become a popular paradigm where services and applications are deployed at the network edge closer to the data sources. It provides applications with outstanding benefits, including reduced response latency and enhanced privacy protection. For emerging advanced applications, such as autonomous vehicles, industrial IoT, and metaverse, further research is needed. This is because such applications demand ultra-low latency, hyper-connectivity, and dynamic and reliable service provision, while existing approaches are inadequate to address the new challenges. Hence, we envision that the future edge computing is moving towards distributed intelligence, where heterogeneous edge nodes collaborate to provide services in large-scale and geo-distributed edge infrastructure. We thereby propose Edge-as-a-Service (EaaS) to enable distributed intelligence. EaaS jointly manages large-scale cross-node edge resources and facilitates edge autonomy, edge-to-edge collaboration, and resource elasticity. These features enable flexible deployment of services and ubiquitous computation and intelligence. We first give an overview of existing edge computing studies and discuss their limitations to articulate the motivation for proposing EaaS. Then, we describe the details of EaaS, including the physical architecture, proposed software framework, and benefits of EaaS. Various application scenarios, such as real-time video surveillance, smart building, and metaverse, are presented to illustrate the significance and potential of EaaS. Finally, we discuss several challenging issues of EaaS to inspire more research towards this new edge computing framework.
\end{abstract}


\section{Introduction}
%Recently, there has been a noticeable shift to migrate the computation-intensive workloads from the remote cloud to near-end edges \cite{shi2016edge}. Compared with traditional cloud computing, the emerging edge computing paradigm enjoys outstanding benefits, including reduced response latency and enhanced privacy preservation \cite{mao2017survey}\cite{chen2019deep}. Edge computing brings intelligence closer to the network edge, even embedding it into IoT devices accessible by the end users. 

The cloud computing paradigm is a service provisioning model that provides user access to scalable computation, networking, and storage resources in the cloud data centers \cite{qian2009cloud}. Cloud service providers provide flexibility and efficiency for end-users by providing services such as software as service (SaaS), platform as a service (PaaS), and infrastructure as a service (IaaS) \cite{kavis2014architecting}. Cloud computing has tremendously changed how we live, work, and study and has become a mainstream business model for many organizations. 

However, cloud computing shows significant disadvantages with the burgeoning of the Internet of Everything. According to Cisco, 80\% of data will be generated by massive IoT devices at the network edge by 2025. Processing such large amounts of data in the cloud will lead to high latency and bandwidth costs, as well as privacy concerns. Recently, edge computing is emerging as a promising paradigm that processes data and provides service on edge devices closer to the data sources \cite{shi2016edge}. Edge computing brings intelligence to the network edge, even embedding it into IoT devices accessible by the end-users.

In the past few years, many research issues have been studied to deploy applications and services at the edge, such as computation offloading \cite{mao2017joint}, content caching \cite{safavat2020recent}, and edge AI \cite{wang2020convergence}. However, existing works are inadequate to meet the requirements of emerging advanced application, such as autonomous vehicles \cite{liu2019edge}, industrial IoT \cite{qiu2020edge}, and metaverse \cite{duan2021metaverse}. This is because those applications demand ultra-low latency, hyper-connectivity, dynamic and reliable service provision, and the support of distributed and collaborative edge computing. More specifically, most existing works focus on edge-cloud collaboration to share data and computation resources while neglecting the collaboration among edge devices. They suffer from the unpredictable latency between the cloud and edge devices and incur privacy concerns. Moreover, they assume a hierarchical and centralized cloud-edge-end architecture, where the cloud dominates and manages the applications and resources of both edge and end devices. However, such a centralized architecture is with poor scalability in achieving large-scale deployment and resource elasticity.    

To support those advanced applications and provide ubiquitous intelligence, we envision that with wider availability of edge computing infrastructures, future edge computing is moving towards collaborative distributed intelligence, where heterogeneous edge nodes collaborate with each other in large-scale and geo-distributed edge computing environments for various computation workloads, including collaborative task processing, distributed machine learning .etc. 

To enable distributed intelligence, we propose a novel edge computing framework, namely Edge as a Service (EaaS), to facilitate edge autonomy, edge-to-edge collaboration, and large-scale application deployment and resource elasticity. Specifically, EaaS jointly manages the large-scale, geo-distributed, and heterogeneous edge infrastructures to boost the collaboration among edge devices, which enables the flexible and large-scale deployment of services, making computation and intelligence ubiquitous. EaaS also provides holistic and end-to-end services for developing and deploying applications at the edge, including IaaS, PaaS, and SaaS. 

Some works also consider the idea of edge computing as a service. Davy et al.~\cite{davy2014challenges} proposed the concept of edge-as-a-service, which aims to leverage the network virtualization to facilitate the flexible usage of access network infrastructure and generate revenue for network operators. Varghese et al.~\cite{varghese2017edge} presented an EaaS platform with a node discovery protocol to enable the selection and deployment of applications on edge. Differently, our work considers edge-as-a-service from a broader perspective. It is a cross-layer service-oriented computing framework integrating large-scale and geo-distributed resource management, edge-native task scheduling, various service modules, and application development throughout the infrastructure, platform, and application layers.

The main contributions of this work are as follows.
\begin{itemize}
    \item We envision that the future edge computing is moving towards distributed intelligence, where edge nodes collaborate with each other to support various applications in large-scale, heterogeneous, and geo-distributed edge infrastructure. 
    \item We propose EaaS, a novel service-oriented edge computing framework, to enable future edge computing by facilitating edge autonomy, edge-to-collaboration, and large-scale service deployment.
    \item Motivations, software framework, benefits, and potential application scenarios of EaaS are discussed.
    \item We present several open issues and challenges to inspire more research towards distributed intelligence empowered by edge computing.
\end{itemize}

The rest of the paper is organized as follows. Sec.~\ref{sec:motivation} discusses existing edge computing research and the motivation for proposing EaaS. Sec.~\ref{sec:framework} introduces in detail about the software framework and the benefits of EaaS. In Sec.~\ref{sec:application}, we show the potential application scenarios of EaaS, including real-time video surveillance, smart building, autonomous vehicles, and metaverse. Sec.~\ref{sec:challenge} discusses the open issues and challenges associated with EaaS. In Sec.~\ref{sec:conclusion}, we conclude this work.

\section{Motivation for future edge computing} \label{sec:motivation}
In this section, we first introduce the existing research and systems of edge computing and then describe the limitations of existing solutions to meet the requirements of future edge computing applications. Finally, we depict the vision and key features of future edge computing.  

\subsection{Introduction to Current Edge Computing}
Edge computing has attracted lots of attention from both academics and industry in recent years. The different works in literature have studied several problems with respect to computation offloading, edge AI, designing applications for different domains, and developing edge computing systems.

Computation offloading, also known as task offloading, aims to offload the computation tasks from end devices to edge devices or the cloud. \cite{zhao2019computation,lyu2017optimal,mao2017joint} offload computing-intensive tasks of end devise to an edge server for processing with the objective of minimizing the task completion time. However, the edge device may easily get overloaded and lead to degraded performance considering the limited edge resources. It is beneficial to incorporate the cloud into the offloading decision. \cite{chen2017engine,meng2019dedas,han2019ondisc} jointly considers the computation resources and networking capacities of the cloud, edge devices, and user devices to explore the possible offloading options. Nevertheless, the collaboration between cloud and edge devices suffers from unpredictable latency and causes privacy issues. 

Another problem is edge AI. Due to the recent advances in deep learning, various AI applications are emerging to provide advanced analytical functions, such as object detection and tracking and human action recognition. The edge AI aims to facilitate the collaborative inference and training of AI models among cloud, edge, and end. Li et al. \cite{li2019edge} model the execution delay and energy consumption of each layer of a neural network model to determine the partition point and distribute the neural network between end and edge, thus accelerating the DNN inference. Ding et al. \cite{ding2020cloud} first used historical data to train the model in the cloud and then shared the shallow network layer EdgeCNN to the edge. Later, the edge used the newly collected real-time data to optimize the model to reduce transmission delay and improve user experience.   

Apart from the existing research work, there are also some edge computing systems. Many start-ups and industry leaders in cloud computing, such as Microsoft, Amazon, and Huawei, have developed edge computing platforms to develop and deploy applications on resource-constraint devices, including AWS IoT Greengrass \cite{kurniawan2018learning}, Azure IoT Edge \cite{jensen2019beginning}, and KubeEdge \cite{xiong2018extend}. However, most of these solutions aim to extend the cloud computing functionalities. These solutions often follow the principles of service-oriented architecture to deploy cloud-based services on edge devices that are close to data sources. Table.~\ref{t:comparison} shows a comparison between the existing commercial and open-source edge computing systems. We can see that exiting edge computing provides less support to edge autonomy and edge-to-edge collaboration, as they rely on centralized management from the cloud.

\begin{table*}[t]
    \centering
    \caption{Comparison of existing edge computing systems.}
    \label{t:comparison} 
     \begin{tabular}{cccccc}
        \toprule
        System & Service-oriented & Programmable & Resource Elasticity & Edge Autonomy & Edge-to-edge Collaboration\\
        \toprule
         IoT Greengrass & yes & yes & supported & supported & limited \\
        Azure IoT Edge & yes & yes & supported & limited & limited \\ 
        Cloud IoT Edge & yes & yes & supported & limited & limited \\
        KubeEdge & yes & yes & supported & supported & limited \\
        Baetyl & yes & yes & supported & supported & limited\\
        OpenYurt & yes & yes & supported & supported & limited \\
        Edge Foundry & yes & limited & supported & limited & limited\\
        Edgent & yes & yes & supported & limited & limited\\
        \bottomrule
     \end{tabular}
\end{table*}

\subsection{Increasing Demanding Application Requirements}
%Although lots of edge computing framework has been proposed. Most of them are designed for existing application scenarios and operate as an extension for the cloud. Meanwhile, the dramatic development of information and communication technologies has brought striking revolutions to tremendous disciplines and various new requirements emerges. Ever increasing number of smart devices have been manufactured and used both in individual and public areas. 5G techniques make it possible for everyone to access to stable and high-bandwidth communication. The edge devices are also supposed to provide dynamic access for mobile end devices carried by users. In the vision of future smart city, the edge device is an essential component in the whole system and it should be able to support computation intensive application like AI algorithms for various application scenarios such as smart factory, autonomous vehicle, smart home and AR/VR. The aforementioned three types of requirements are pushing the developments and shaping the future of edge computing.

Though achieving great success in the past few years, current edge computing is inadequate to serve the emerging advanced applications. Specifically, the dramatic development of information and communication technologies has brought striking revolutions to tremendous disciplines, and various new applications have emerged, including autonomous vehicles, industrial IoT, and metaverse. Those advanced applications are expected to be deployed at the network edge and pose great challenges to current edge computing. Those challenges are summarized as follows.
\subsubsection{Ultra-low latency} To achieve safe autonomous vehicles and immersive interactive experience in metaverse, the edge devices are required to perform computation-intensive tasks in milliseconds, such as object detection and tracking, human action recognition, and 3D point cloud reconstruction. Existing works offloading the computation tasks to the cloud may suffer from the unpredictable latency between the cloud and the edge devices and leads to privacy concerns.

\subsubsection{Large-scale deployment with hyper-connectivity} For industrial IoT and metaverse, tremendous end devices such as VR glasses, industrial controllers, and sensors are required to be interconnected for online interaction and data collecting. Existing edge computing approaches usually adopt centralized management, which is with limited scalability and fails to support the hyper-connectivity of end devices. Further, those end devices are usually located in different geographical areas and require collaboration among geo-distributed edge devices to provide large-scale service deployment.

\subsubsection{Dynamic and reliable service provision} Many end devices are with high mobility, such as mobile phones and autonomous vehicles. They frequently pass through distinct geographical areas and demand consistent and reliable service provision. Existing works rely on the centralized cloud to synchronize the service status and make the service migration decisions, which is vulnerable to the dynamic network and is insufficient to provide reliable service provision. The dynamic network connection and service provision require collaboration among geo-distributed edge devices to share both data and computation resources.

%Except for the requirements from future applications, the development of related technology also play a decisive role in the developing of future edge computing. Similar to cloud computing, the key technology in edge computing can be classified into two categories, software and hardware. The hardware developing trend in recent years basically follows the update of computation devices (GPU, FPGA, TPU, and NPU) and storage devices such as SSD. Virtualization technology is the essential tech for both cloud and edge computing because it vitalizes the resources from heterogeneous hardware. Services can run on multiple devices which is the foundation of distributed system. It is widely accepted that virtualization technology like virtual machine and container are still the foundation of edge computing in the near future.

\begin{figure}[t]
    \centering 
    \includegraphics[width=\linewidth]{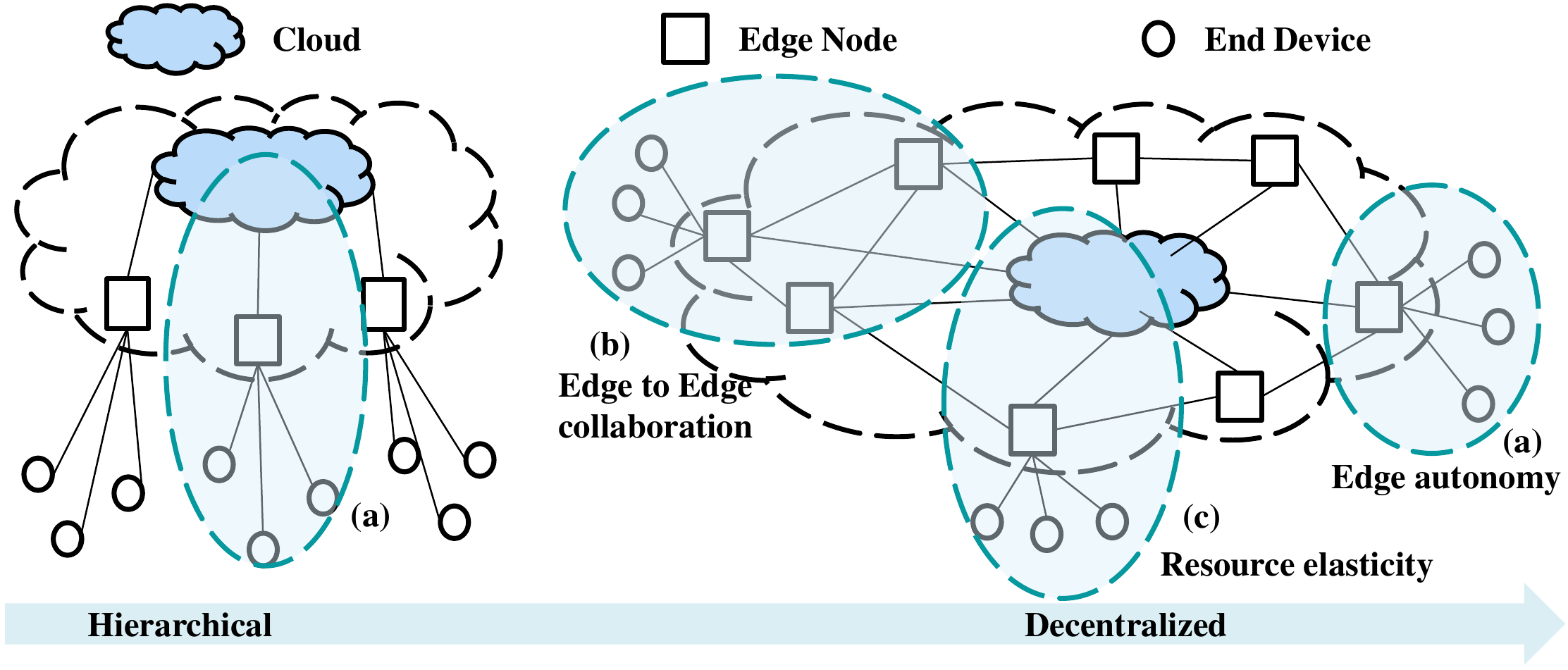}
    \caption{Architecture of Future Edge Computing.}
    \label{f:fig_sim} 
\end{figure}

\begin{figure*}[t]                
    \centering 
    \includegraphics[width=.8\linewidth]{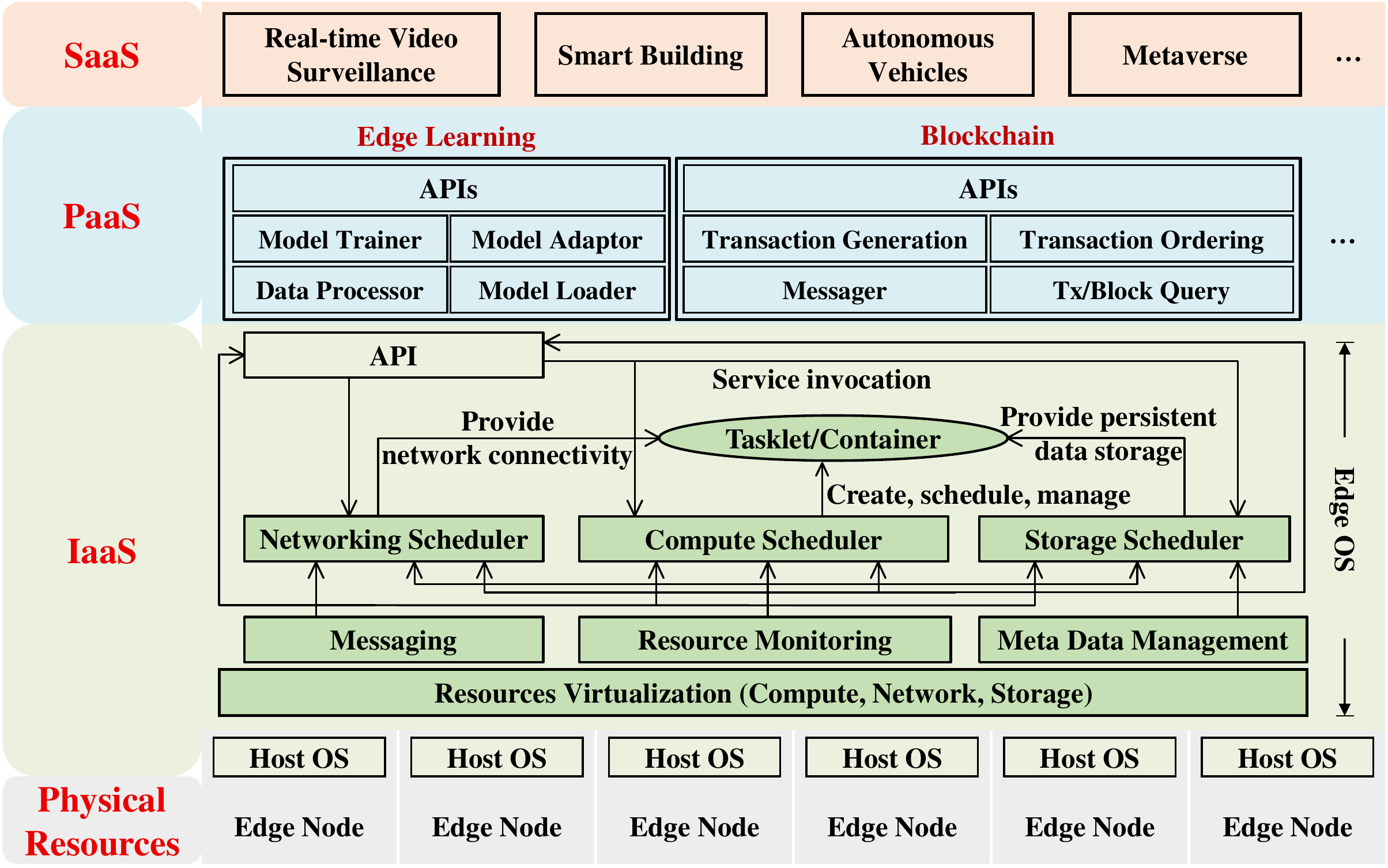}
    \caption{Framework of Edge as a Service.}
    \label{f:framework} 
\end{figure*}

\subsection{Vision of Future Edge Computing}
To better support emerging advanced new applications and provide ultra-low latency, reliable, and ubiquitous intelligent services, further research to enhance current edge computing is needed. Hence, we envision that the future edge computing is moving towards distributed intelligence, where numerous edge devices collaborate with each other in the large-scale, heterogeneous, and geo-distributed edge infrastructure.    
 
There are three new key properties of future edge computing to enable distributed intelligence, i.e., edge-to-edge collaboration, edge autonomy, and resource elasticity. The three properties  can be described as:
\begin{itemize}
    \item Edge autonomy: It implies enabling offline and/or localized operations for reliable service provision without communication with the cloud.
    \item Edge-to-edge collaboration: It implies that each edge device can collaborate with its surrounding edge nodes to form a cluster that has the ability to provide real-time processing and computation-intensive services like AI algorithms. Furthermore, seamless and cross-platform computing can also be accomplished among edge nodes, and dynamic access can be subsequently supported.
    \item Resource elasticity: It refers to the ability of a system to support autonomous scheduling, including provisioning and de-provisioning, of different resources and services for a large-scale, distributed, and heterogeneous edge computing environment. 
\end{itemize}

These three properties are essential to support the different requirements of future edge applications, including large-scale deployment, real-time processing, and dynamic access and reliable service provisioning. The existing solutions are insufficient to support these stringent requirements of future edge applications. Furthermore, The existing platforms currently lack in providing an integrated solution for future edge computing that includes providing infrastructure, platform, and software level services for the end-to-end development of edge computing applications. There is a need for a paradigm shift, as shown in Fig.~\ref{f:fig_sim},  from the hierarchical architecture used in existing platforms to decentralized architecture for future edge applications. In future edge computing, the cloud still exists but will not play the dominant role in managing all the edge and end nodes.

\section{EaaS Framework} \label{sec:framework}
In this section, we give details about the EaaS framework, which aims to enable collaborative distributed intelligence supporting various advanced applications. We first give an overview of the EaaS software framework and then articulate the benefits of EaaS.

\subsection{EaaS Framework}
EaaS is designed to manage the large-scale, geo-distributed, and heterogeneous edge resources and support the collaboration among edge nodes for various emerging advanced applications. It mainly consists of three layers, i.e., infrastructure as a service (IaaS), platform as a service (PaaS), and software as a service (SaaS), as shown in Fig.~\ref{f:framework}. 

\subsubsection{Infrastructure as a Service} In edge computing scenarios, there are various distinct edge nodes, such as edge servers, edge gateways, and intelligent vehicles, which are with heterogeneous computation, storage, and networking capabilities. Services and applications are deployed on those geo-distributed nodes connected with low-bandwidth and intermittent networks. 

A fundamental problem of IaaS is to manage the large-scale, heterogeneous, and geo-distributed edge resources and schedule application workloads among edge nodes for efficient task execution. Different from the resource management and workload scheduling in the cloud environment, where cloud servers with abundant computation resources are connected with a high-bandwidth and stable network in a data center, resource management and workload scheduling in edge computing is more challenging. 

First, edge-native applications are usually performance-aware, demanding high throughput and low latency. The workload scheduling in the cloud environment is mainly to ensure resource provision of workloads, such as the capacity of requested memory and CPU cores. It lacks support to meet the performance requirements of edge-native applications. Second, edge-native applications are with inner dependencies. Many intelligent edge applications are resource-greedy and complex, consisting of lots of inter-dependent components which are usually deployed to multiple edge nodes considering the constraint resource of a single node. However, the workload scheduling in a cloud environment usually schedules the application to a server with abundant resources and fails to consider the application's inner structure. Third, the data, computation, and networking resources are heterogeneous and coupled with each other. Application deployed on heterogeneous edge nodes experiences distinct performance, and the coupled resources require joint orchestration. However, cloud resource management concentrates on orchestrating computation resources without jointly considering the data locality and networking resources, which may lead to underutilized resources and poor performance of workloads. 

Hence, the new solutions of resource management and workload scheduling need to jointly consider the heterogeneous and coupled data, networking, and computation resources and make optimized scheduling decisions to meet the performance requirements of edge-native applications.
In the infrastructure layer, we design a distributed edge computing operating system, namely EdgeOS, to manage the cross-node coupled edge resources and facilitate efficient workload scheduling among the geo-distributed edge nodes.

To manage the cross-node and coupled edge resources of computation, networking, and storage, EdgeOS adopts lightweight virtualization technology, i.e., containers, to abstract the heterogeneous resources and enable seamless service migration among geo-distributed edge nodes.
Above the abstracted resources, we design three task schedulers, i.e., networking scheduler, compute schedule, and storage scheduler, to manage the networking, compute, and storage resources, respectively. The three schedulers collaborated with each other to decide and maintain the resource allocation strategies for the incoming computation tasks. Moreover, EdgeOS provides APIs for the platform services to easily call the functions.

\begin{figure}[t]                
    \centering 
    \includegraphics[width=.9\linewidth]{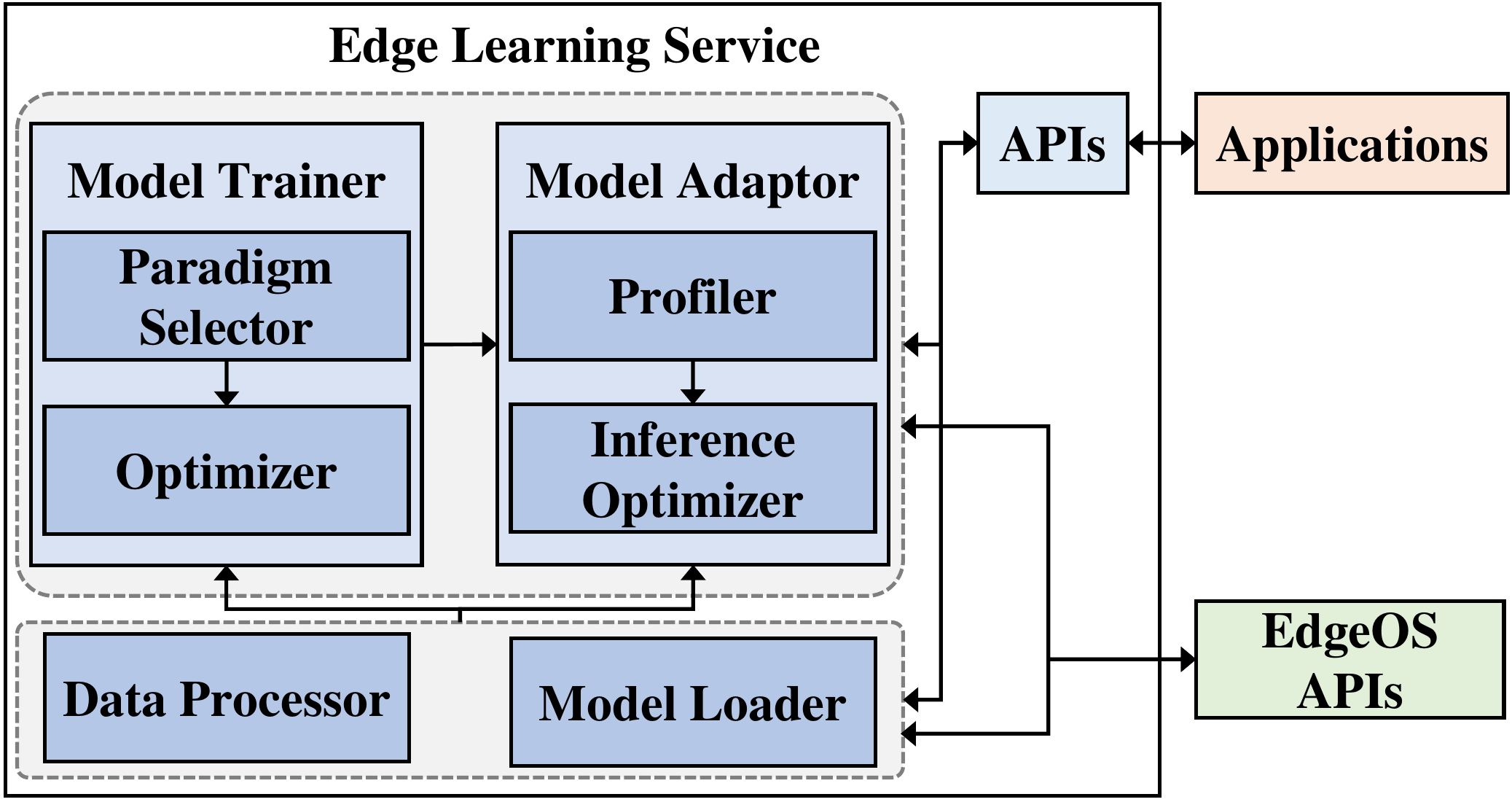} 
    \caption{Framework of Edge Learning Service.}
    \label{f:edge_learning} 
\end{figure}

\subsubsection{Platform as a Service} Above the edge infrastructures, many platform-level services, such as edge learning service and blockchain service, can be developed to support various applications.

One important service is edge learning service. Edge learning, also terms as edge AI, refers to the training and inference of AI models near the users at the network edge. Edge AI is essential to support applications requiring short communication delays and high response speed. There are many edge AI applications deployed at the network edge. Edge learning service is designed to support the whole life-cycle of AI applications, including data preparation, algorithm development, model training, and model inference. Edge learning service is in the platform layer. It provides APIs for easy and fast application development in the upper layer and calls the APIs provided by the infrastructure management layer to achieve efficient application deployment and resource utilization over heterogeneous and geo-distributed edge nodes. 

The functional module design of the edge learning service is shown in Fig.~\ref{f:edge_learning}. There are mainly four modules of edge learning service, namely, Data Processor, Model Loader, Model Trainer, and Model Adaptor. The data processor is responsible for data preprocessing, including data sampling, extracting, and transformation. It also supports data self-labeling and human labeling. The model loader supports algorithm development for users. It provides various programming libraries for AI model development and light-weighted pre-trained models, such as Resnet-50 and AlexNet. The model trainer is designed for conducting distributed model training. It provides multiple training paradigms, including federated learning, gossip learning \cite{hegedHus2019gossip}, and E-tree learning \cite{yang2021tree}. Further, an optimizer is designed to support automatic parameters optimization of distributed model training, such as batch size, learning rate, and model aggregation frequency. With intelligent optimization algorithms, the distributed model training can adapt to the dynamic resources of the edge environments and achieve improved accuracy and convergence speed. The model adaptor is mainly responsible for optimizing the model inference strategies. It adopts various optimization methods, including model compression \cite{courbariaux2015binaryconnect}, model partition \cite{kang2017neurosurgeon}, and model early exit \cite{li2018learning}, to enable resource-aware model inference to meet the applications’ requirements, such as latency, energy, and fault tolerance.   

We take the model inference as an example to show the workflow of the edge learning service. The model adaptor takes the model inference request as input and generates the execution strategies. Specifically, the inference request defines the data sources, the AI models, and the performance requirements, such as throughput and latency. The profiler first profiles the execution time of AI models on heterogeneous edge nodes and generates the profile information. The profile information and the underlying resource status accessed from EdgeOS APIs, as well as the inference optimization techniques, including model compression, partition, and early exit, will be jointly considered to decide the inference strategies. The inference strategies define how to compress the model, or how to partition the model, and where to allocate the inference tasks so as to meet the performance requirements. The strategies will be executed by the EdgeOS to achieve efficient model inference.

Edge blockchain as a service (EBaaS) is another essential service besides edge learning \cite{jindal2019survivor}. In EBaaS, the blockchain components are deployed at edge devices near the users and provided as easy-to-use services \cite{tuli2019fogbus}. In this manner, the blockchain services can be delivered to the users with great convenience and low latency. The ultimate goals of EBaaS are delivering blockchain services with high modularity, flexibility, scalability, reliability, and security \cite{jiang2021polychain}.

The current development of blockchain technology needs to go through two stages towards EBaaS: service-oriented and edge deployment. In the current stage, the companies are developing blockchain applications and hosting blockchain platforms by themselves \cite{jiang2020fairness,singh2018blockchain}. They analyze the specific requirements of different applications, e.g., privacy and efficiency \cite{jiang2019privacy}, design variant blockchain solutions, develop heterogeneous blockchain components, and deploy the blockchain applications on separated hardware. Although the applications share similar requirements and blockchain components, the companies rarely considered integrating the different applications in a unified blockchain infrastructure, e.g., cloud and edge computing environments. Note that there are some popular blockchain platforms, e.g., Bitcoin \cite{nakamoto2008bitcoin}, Ethereum \cite{kim2018measuring}, and Hyperledger Fabric \cite{androulaki2018hyperledger}. However, they are provided as blockchain solutions rather than services. The users still must adapt them for different applications and host the blockchain components.

More recently, there is emerging research on blockchain as a service, aiming at developing a service-oriented blockchain infrastructure that can be used conveniently \cite{zheng2019nutbaas,lu2019ubaas,chen2018fbaas, mamun2021baash,jiang2021polychain}. For example, Jiang et al. developed PolyChain, a generic blockchain as a service platform \cite{jiang2021polychain}. They decompose a logical blockchain node into four components: application, consensus, storage, and network. Then, the components of a logical blockchain are deployed on different physical machines considering the resource demand of components and supply from machines. The techniques are named component modularization, distributed deployment, and resource optimization. With blockchain as a service, a generic blockchain infrastructure can be set, on which numerous applications can be deployed \cite{aujla2020blocksdn}.

Traditionally, the blockchain as a service infrastructure is deployed on the cloud, and the blockchain services become cloud-based \cite{weerasinghe2021novel}. However, the cloud is far from the end-users, limiting the application performance \cite{tsai2010service}. With the technological development of edge computing, EBaaS comes out to supply blockchain services near the end-users. We believe EBaaS will be favorable and trendy to support the applications demanding high security, privacy preservation, great flexibility, and low latency, which are not feasible currently.

\subsubsection{Software as a Service} SaaS is defined as an application hosted by cloud servers to offer specific services such as email (Gmail) and storage (Dropbox) via a subscription-based licensing model. EaaS enables SaaS providers to deliver faster, privacy-preserving services for their customers by leveraging the outstanding benefits of edge computing. However, EaaS also poses new challenges to the way that SaaS providers develop their applications and provide services. 

First, the application should adopt a modular and stateless design, which decomposes the functions of the applications into a set of dependent modules. Considering the limited resources of a single edge node, deploying the whole application on an edge node may be infeasible. Hence, the application is required to be partitioned and deployed on multiple edge nodes. Dataflow programming model \cite{johnston2004advances} could be one option to enable efficient modular application design. Further, the frequent application state migration may lead to network congestion and degrade the applications. Stateless programming is beneficial in developing SaaS applications.

Second, the application development is dependent on the underlying PaaS and IaaS. Specifically, edge nodes are usually heterogeneous with different architectures, such as x86, ARM. Some edge nodes may be equipped with TPU, FPGA. Considering all the hardware characteristics may be challenging for developers, while the platform services may abstract the heterogeneity and provide unified programming interfaces. Similarly, many applications incorporate 5G capability into their functions, which requires the networking programming abstractions from the IaaS. 

\begin{figure}[t]                
    \centering 
    \includegraphics[width=0.9\linewidth]{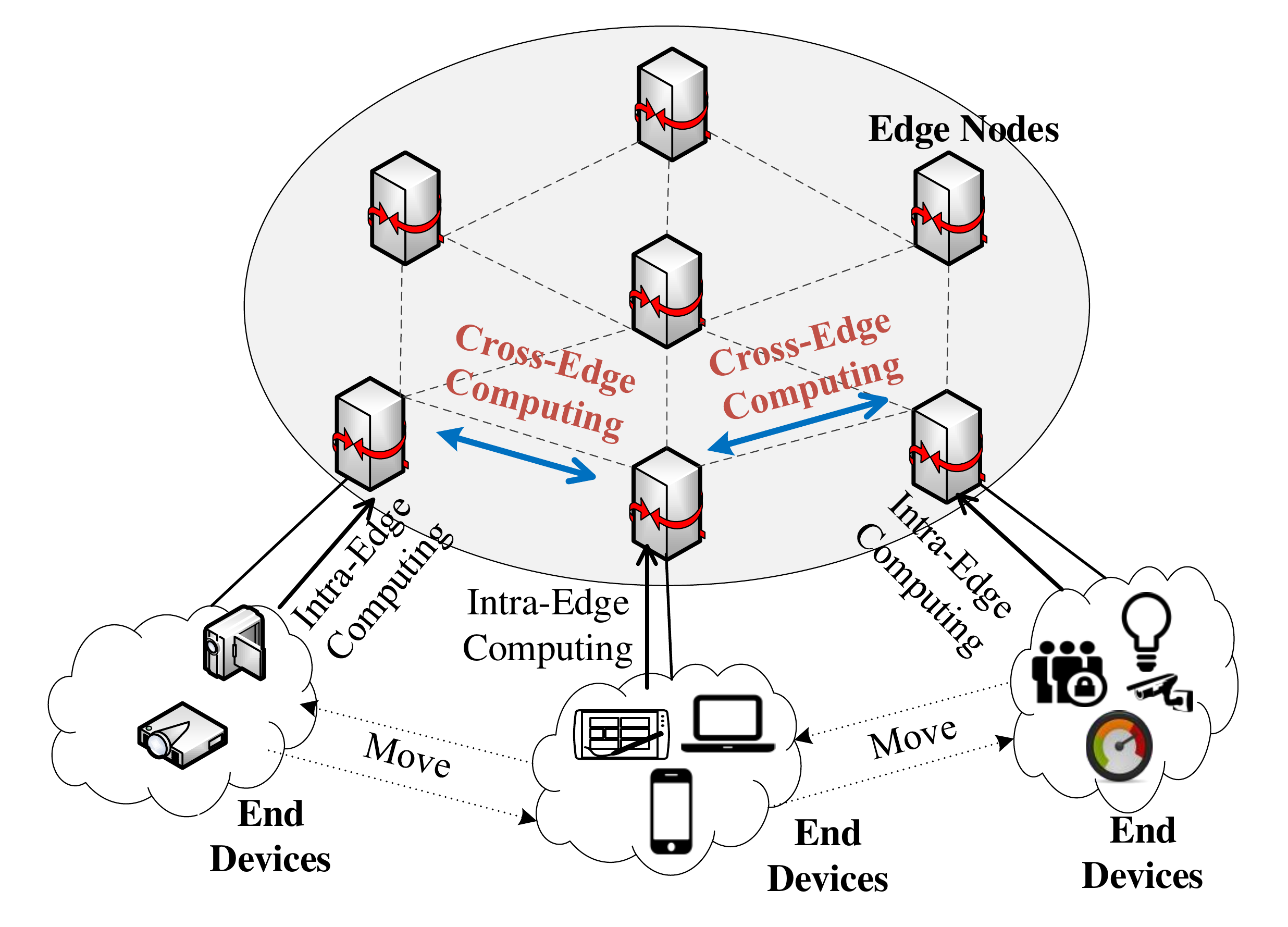} 
    \caption{Hybrid Architecture of EaaS.}
    \label{f:hybrid_architecture} 
\end{figure}

\subsection{Benefits of the EaaS}
Compared with existing edge computing frameworks, EaaS has outstanding benefits and significance as follows.

\textit{Large-scale connection and deployment.} Current edge computing usually adopts the hierarchical cloud-edge-end architecture. It is a centralized architecture with the cloud to manage the whole system, which suffers from the limited scalability while many applications require large-scale deployment. To achieve the goal of pervasive intelligence, EaaS adopts a hybrid architecture based on our previous proposed Edge Mesh framework \cite{sahni2017edge}, integrating both centralized and decentralized resource management manner. As shown in Fig.~\ref{f:hybrid_architecture}, the edge node manages the end devices in a centralized manner, while edge nodes collaborate with each other in a decentralized manner considering the unpredictable latency between cloud and edge nodes. The hybrid architecture enables large-scale connections between edge nodes and end devices and among edge nodes. 

\textit{Coupled resource management.} In collaborative edge computing environments, data locate on geo-distributed and heterogeneous edge nodes. Considering the limited resources of a single edge node, the computation-intensive applications are more likely to be partitioned and deployed in distributed edge nodes, which may lead to frequent communication through a low-bandwidth network. Hence, the data, computation, and networking resources are coupled with each other and need joint management and orchestration \cite{sahni2018data,sahni2020multihop}. This is beneficial to improve resource utilization and meet the performance requirements of edge-native applications. Most previous works only consider orchestrating computation resources but neglect the resource heterogeneity and network resources, such as bandwidth allocation and customized routing of data flows. We envision that future edge computing is required to jointly orchestrate the heterogeneous computation and networking resources with the consideration of the data locality. 

\textit{Edge-native application development and deployment.} Different from cloud-native applications, edge-native applications are usually performance-sensitive and with inner structures, where dependent application modules are deployed on multiple edge nodes due to the limited resources of a single edge node. Further, edge nodes are with heterogeneous computation resources. Existing approaches concentrate on the development of applications while neglecting the deployment issue of applications \cite{satyanarayanan2019seminal}. They focus less on resource heterogeneity, locality of source data, and inter-communications of edge nodes. Those factors affect a lot of the performance of edge-native applications. Instead, EaaS supports end-to-end support of the application development and deployment. First, EaaS provides universal programming abstractions that enables decomposing the application and explicitly declaring the dependencies of the inner modules. Second, EaaS provides coupled resource management, enabling application profiling on heterogeneous edge nodes and scheduling application workloads for improved distributed execution by jointly considering the couped data, computation, and networking resources. 

\textit{Flexible service deployment.} The traditional cloud-edge-end three-tier architecture is extensively studied in the literature. However, this hierarchical framework provides less support for edge autonomy, flexible service access, and edge-edge collaboration. In the hierarchical framework, the decision-making for task offloading and data transmission are made on the cloud, which monitors and manages the edge and end nodes. When there is an unstable network connection between cloud and edge, the system shows apparent performance degradation and even cannot work properly. Further, for mobile devices, which usually move among geographical areas, the edge nodes may not be able to provide seamless services. Moreover, the existing framework adopts cloud-edge collaboration, which suffers from the unpredictable latency between cloud and edge and may cause privacy issues due to cloud-edge data transmission. Instead, EaaS provides both cloud-edge and edge-edge collaboration to share both data and computation resources, which enables flexible service access and lower latency. When the cloud is unavailable, the edge nodes can make decisions autonomously by collaborating with other edge nodes. EaaS achieves pervasive and ubiquitous intelligence through flexible server deployment and assessment.

\section{Application Scenarios of EaaS} \label{sec:application}

This section discusses the promising applications of EaaS, especially the emerging ones.

\subsection{Real-time Video Surveillance}
Real-time video surveillance plays a crucial role in various areas in a smart city, including automatic pilot, smart transportation, and public security monitoring. Enormous AI-based video surveillance algorithms have been developed for object tracking, object re-identification, action identification, risk behavior detection, and many more. Edge computing is taken as the promising computing diagram since it can support low-latency, proactive, and secure video surveillance by providing computation near cameras. Most existing edge-based solutions utilize edge-cloud collaboration \cite{zhao2020lightweight,chen2020binarized} for the offloading of computation-intensive tasks, such as pedestrian re-identification and risky behavior identification. However, the use of a remote cloud server will degrade the latency and data security. Some studies focus on edge-to-edge cooperation for providing surveillance services jointly \cite{zeng2020distream,xu2019space}.
However, they didn't schedule application workloads by jointly considering the coupled resources, such as locality of cameras (data sources), resource heterogeneity and communication overhead among geo-distributed edge nodes.

Our proposed EaaS represents a promising way of bringing together the strength of the aforementioned edge-based solutions, as it supports both edge-to-edge collaboration and resource elasticity while considering the coupled resources. It provides the flexibility to invoke nearby edge nodes and remote cloud servers efficiently for adapting to the ever-change computation requirements in various video surveillance scenarios. For instance,  edge-to-edge collaboration is beneficial for proactive pedestrian re-identification in medium-range areas such as university campuses, compared with edge-cloud collaboration. This is because data (such as pedestrian features) and computation resources sharing among nearby edge devices provides lower latency for reconstructing the pedestrian's trajectory. Instead, suppose the application is to perform pedestrian re-identification on a city scale. Offloading the re-identification task to the cloud is preferable as searching for suspects from numerous surveillance cameras in a city is computation-intensive and requires significant storage space.

\subsection{Smart Building}
As a crucial AIoT application, smart building plays a fundamental role in the safe and efficient production of society. It promises the safe and efficient operation of engineering structures, including bridges, high-rise buildings, tunnels, railways, power plants, etc. A smart building consists of multiple subsystems such as structural health monitoring system, drainage condition monitoring system, and power monitoring system. The conditions of each subsystem are identified from the measurement data collected by sensors, including accelerometers, strain gauges, cameras, displacement sensors, GPS, temperature sensors, anemometers, etc \cite{sahni2018midshm}. A cloud-based smart building system, as the mainstream solution, can support elastic computation and data storage resources for smart building applications. But they are inapplicable for data-sensitive and latency-sensitive smart building applications. Also, the collaboration between subsystems is dragged down by the centralized cloud server. Therefore, edge computing was introduced to smart building and has been extensively studied recently. Nevertheless, most existing studies focus on optimizing smart building algorithms for edge devices \cite{chen2022physics, wu2019pruning}. Works to optimize the edge computing system are very limited. Our proposed EaaS can work as the underlying system for the dedicated smart building applications designed for distributed edge devices. Specifically, EaaS enables efficient communication between subsystems in a smart building through edge-to-edge collaboration. Also, it supports efficient execution of distributed data processing algorithms such as modal analysis \cite{liu2014distributed}, multivariate correlation analysis \cite{yu2021review}, and anomaly detection \cite{sater2021federated} among multiple edge devices.

\subsection{Autonomous Vehicles}
An autonomous vehicle is a vehicle that can sense its surroundings and operate without human intervention. The autonomous vehicles tightly integrated many technologies, including sensing, localization, perception, decision-making, and mechanical control \cite{liu2019edge}. The primary task of autonomous vehicles is to realize automated decision-making of the running paths, which includes AI computation tasks such as identification and tracking of vehicles and pedestrians, trajectory prediction, and path planning. The decision-making of running path task has demanding requirements of the execution latency, usually less than 5ms. However, due to the limited on-board computation power of a single vehicle, it is essential to offload the running path planning task to other computation devices. Recently, the vehicular edge computing network has been proposed to enable communication and resource sharing among autonomous vehicles and nearby edge nodes, such as base stations and road-side units \cite{xie2019collaborative}. Most existing work considers offloading the computation task to a nearby base station or a road-side unit. However, the connection between vehicles and base stations maybe not always available, and the base station may get overloaded if all tasks are offloaded to a base station \cite{cui2020offloading,yang2019efficient}. Instead, EaaS jointly manage the resources of base stations, road-side units, and vehicles and enable efficient task offloading among all available edge nodes. It leads to better resource utilization and reduced service latency and is more suitable in unstable network connection.

\subsection{Metaverse}
Metaverse is a digital 3D world where people can work, shop, play, socialize, and do many more things they can do in real life. It is regarded as a promising future direction of the Internet and has attracted numerous investments from tech giants. The main objective of metaverse is realizing the physical experience for users in a digital world. To provide users with highly immersive experiences, a high-resolution, interactive, and low-latency digital world is desperately needed. Nevertheless, building such a digital world requires enormous computational and storage resources for the execution of physics calculation, rendering, and AI models \cite{dhelim2022edge}. Edge computing has been regarded as a crucial technique for promising the efficient execution of those applications due to its nature of low latency. But the limited computation resources in user's edge devices (such as VR/AR devices, mobile devices, and personal computers) hinder the development of edge-empowered metaverse \cite{xu2021wireless,kim2021edge}. Moreover, the large-scale deployed edge devices are heterogeneous and might not be able to provide the same user experience for each people in the metaverse. EaaS provides an integrated solution for running metaverse applications near users. On the one hand, EaaS supports flexible service deployment. The computation workload can be taken by a single edge device in edge autonomy mode, shared with nearby MEC servers through edge-to-edge collaboration, or offloaded to remote cloud servers. The modes are changed automatically regarding the variation of workload, the resources in local edge devices, and available nearby edge resources. On the other hand, EaaS has incorporated multiple native edge learning services for efficient execution of AI models even with limited resources.

\section{Open Challenges} \label{sec:challenge}
This section describes the open research challenges that have not been fully explored while designing and developing the services in EaaS.

\subsection{Lightweight and Cross-platform Virtualization}

Virtualization is essential for abstracting the heterogeneity of underlying infrastructure resources in developing EaaS. In the past few years, Container has become a popular approach for achieving lightweight virtualization at the edge devices compared to the alternative approach of virtual machines (VM). However, there are still many issues that need to be resolved to meet the demanding requirements of emerging applications. First, there are few comprehensive solutions for achieving virtualization for embedded edge devices with limited memory, storage, and compute capacity. Docker, a popular container virtualization solution, has minimum requirements on the underlying resources of the host device, for example, at least $4GB$ RAM. Hence, it is not suitable for many resource-constraint embedded edge devices. One trend for virtualization in the future would be to develop lightweight virtualization solutions that can support resource-constraint embedded edge devices. The second issue is supporting cross-platform application deployment, including different processing architectures, i.e., x86, ARM, etc., different types of OS, i.e., GPOS and RTOS, and different processing units, i.e., CPU, GPU, TPU, etc. The current state-of-the-art Docker solution does not perform well for cross-platform application deployment consisting of heterogeneous devices with both Windows and Linux processing architectures and different processing units. Docker also does not officially support RTOS that is required for the deployment of applications with real-time requirements on embedded systems. The third issue is supporting multi-tenancy, i.e., support management and deployment of multiple application services simultaneously. There have been some solutions by Kubernetes to support multi-tenancy, but these solutions make some compromises on the security while deploying multiple application services. One future direction is balancing the trade-off between multi-tenancy and security. 

\subsection{Edge-native Resource Scheduling}

Resource scheduling is a fundamental problem that has been widely studied in the context of edge computing and other related computing paradigms. Many works have studied different scheduling problems such as computation offloading, data caching, bandwidth allocation, energy scheduling, service migration, etc., under different system models and objective functions. However, these works have not fully explored the different characteristic features of EaaS, which makes the edge-native resource scheduling problem more challenging. One main characteristic feature of EaaS is the coupled resources that require considering dependency among resources to make joint scheduling decisions \cite{sahni2020multi}. However, most existing solutions often provide a solution for scheduling a single resource. Novel cooperative scheduling solutions are required for the efficient management of coupled resources. Besides resource dependency, another characteristic feature of EaaS is data and resource uncertainty due to network dynamics and devices/link failures that makes the scheduling problem more complex. The scheduling solution has to be designed by modeling the reliability of devices and network links. There are few works that have considered reliability-aware resource scheduling for edge computing environments \cite{liu2021reliability,peng2019reliability}. However, these works do not focus on the joint task, network, and storage scheduling of coupled resources considering the reliability constraints. The third characteristic feature of EaaS is a large-scale network with distributed resources that makes it difficult to schedule resources using a centralized controller efficiently. Most existing resource scheduling solutions for edge computing often assume global knowledge of resources that may not be practically feasible. Future works on resource scheduling should focus more on distributed scheduling approaches to enable scalable and high-performance applications.

\subsection{Distributed Data Sharing}

The different edge devices need to share and synchronize stateful data to collaborate efficiently with applications in connected healthcare \cite{jiang2018blochie} and smart logistics \cite{wu2019data}. While the topic of distributed data sharing has been studied in the context of traditional parallel and distributed systems and cloud, It is much more challenging to enable it in the context of edge computing. First, the number of edge devices used for edge computing applications is at a much larger scale compared to the cloud. Furthermore, these devices are heterogeneous, resource-constraint, and highly distributed, which makes it difficult to design suitable scalable yet lightweight solutions. Second, the edge devices are often connected using an unstable wireless network that is dynamic and uncertain. The bandwidth across different links is also heterogeneous. Third, It is difficult to ensure consistency of stateful data as existing consensus algorithms are not suitable for highly distributed, large-scale, and dynamic edge computing environments. Novel lightweight consensus mechanisms are required that can ensure consistency for resource-constraint edge computing environments \cite{wu2018tsar}. Furthermore, the new solutions should also consider addressing different consistency requirements as there could be multiple simultaneously deployed application services with different consistency requirements. 

\subsection{Resource-aware Edge AI}

Edge AI has been identified as a key technology for 5G/6G and future networks by supporting training and deployment of AI models at edge devices close to data sources to enable several benefits, including low-latency processing, higher scalability, and better privacy. Several emerging IoT applications, such as autonomous vehicles, and metaverse, will be enabled by integrating EdgeAI with underlying 5G/6G networks. A characteristic feature of EdgeAI is that data required for training AI models are distributed at multiple edge devices. There are many challenging issues emerging from data properties that need to be resolved to support edge AI. The first issue is model inaccuracy due to the NonIID data generated on the edge devices. Existing edge AI approaches target learning a generalized model fit for all the edge nodes. In practice, data used for training a generalized model differs a lot from the NonIID data generated on an edge node, causing the model inaccuracy for inference on the node. The second issue is data labeling. The data generated on the edge nodes are mostly unlabeled due to high labeling costs. Existing approaches assuming well-labeled data samples cannot work effectively, and new model training methods need to be developed to exploit massive unlabeled data. The third issue is the nature of IoT data which are usually streaming data, continuously arriving at the edge nodes. Existing approaches assuming that data samples are available at the start-up time are not suitable for Edge AI. New approaches are required to train an adaptive and lightweight model at resources constraint edge devices. Besides training AI models, there are also many issues in AI model inference as resource-constraint edge devices have limited compute, storage, and memory capacity to run complex AI models. Therefore, services need to be developed as part of EaaS to adapt the model in real-time by leveraging approaches, such as compression, partition, and early exit, while considering the underlying dependent resources.

\subsection{Security and Privacy}

The different features in edge computing that enable several benefits also lead to some security and privacy concerns. First, the different edge devices are often connected using a wireless network which leads to security and privacy concerns as malicious attackers can access the private data by eavesdropping. Second, as the edge devices are resource-constraint, it is difficult to deploy traditional security and privacy mechanisms that are compute-intensive. Third, the different edge devices can be under the control of multiple stakeholders unwilling to share data leading to privacy concerns. Fourth, the different geographically distributed edge devices are connected and collaborating with each other, which makes it challenging to ensure security and privacy as an attack at one part of the network can spread across other devices. Fifth, the edge devices are often heterogeneous, making it difficult to design generic security and privacy solution suitable for different application and deployment scenarios. Existing works in literature have also identified several new security and privacy issues emerging due to the unique characteristics in edge computing \cite{xiao2019edge,al2022ai,roman2018mobile}.

\subsection{Programming Abstraction}

One major issue to support end-to-end service development and deployment in EaaS is to provide programming interfaces that can abstract the underlying complexities for application developers. Application development and deployment at the edge is much more challenging compared to the cloud environment due to issues such as device/network heterogeneity, geo-distributed resources, network dynamics, resource uncertainty, and stateful data. Programming abstractions should be provided in the form of middleware to allow developers to focus on application logic rather than underlying edge platform issues such as resource discovery, monitoring, and management. Since the edge devices are characterized by limited resources, the middleware should be lightweight with low communication, computation, and storage overhead. The middleware should also be generic, modular, and flexible. The generic requirement implies that middleware should not be application-specific with limited primitives. Instead, it should support the development and deployment of different types of platform services and applications, including edge learning services, blockchain services, and 5G/6G support, for different network and system models. The modular requirement implies that middleware is designed based on the principles of the service-oriented architecture to allow easy replacement of some functionalities and interfaces. Finally, the flexible requirement implies that middleware should be easily extensible to allow the addition of new functions and interfaces.  

\section{Conclusion} \label{sec:conclusion}
Edge computing is a popular computing paradigm where the data processing and intelligence are performed at the network edge closer to the data sources. Many studies have been done in the past few years to deploy applications at the edge to reduce the service response time and bandwidth costs. However, existing approaches are inadequate to support emerging advanced applications demanding ultra-low latency, large-scale deployment with hyper-connectivity, and dynamic and reliable service provision. To address those challenges, we envision that the future of edge computing is moving towards collaborative distributed intelligence, where heterogeneous edge nodes collaborate with each other in a large-scale and geo-distributed edge computing environment. We hence propose EaaS to enable collaborative distributed intelligence by facilitating edge autonomy, edge-to-edge collaboration, large-scale cross-node and coupled resource management, and end-to-end edge-native application development and deployment. At last, we discussed several challenges and open issues to inspire more research toward collaborative distributed intelligence and finally make intelligence ubiquitous in the IoT world.

\section{Acknowledgement}
This work is supported by the Hong Kong RGC General Research Fund under Grant PolyU 15204921 and PolyU 15220922.

\bibliographystyle{IEEEtran}
\bibliography{ref}
\end{document}